\documentclass{article}

 \usepackage[nonatbib, final]{neurips_2020}
 \usepackage[square, numbers]{natbib}

\usepackage[utf8]{inputenc} 
\usepackage[T1]{fontenc}    
\usepackage[colorlinks,allcolors=blue]{hyperref}       
\usepackage{url}            
\usepackage{booktabs}       
\usepackage{amsfonts}       
\usepackage{nicefrac}       
\usepackage{microtype}      

\usepackage{amsmath}
\DeclareMathOperator{\Normal}{\mathrm{Normal}}

\usepackage{algorithm}
\usepackage[noend]{algpseudocode}

\usepackage[table]{xcolor} 
\usepackage{multirow}
\definecolor{LightCyan}{rgb}{0.88,1,1}

\usepackage{tikz}
\usepackage{float}
\usepackage{wrapfig}

\newtheorem{theorem}{Theorem}

\usepackage{listings}
\definecolor{codegreen}{rgb}{0,0.6,0}
\definecolor{codegray}{rgb}{0.5,0.5,0.5}
\definecolor{codepurple}{rgb}{0.58,0,0.82}
\definecolor{backcolour}{rgb}{0.99,0.99,0.97}
\lstdefinestyle{stan}{
	literate={~}{$\sim$}{1},
	backgroundcolor=\color{backcolour},   
	commentstyle=\color{codegreen},
	keywords = {real, vector, matrix, data, model, parameters, transformed, target},
	keywordstyle=\color{magenta},
	numberstyle=\tiny\color{codegray},
	stringstyle=\color{codepurple},
	emph={%
		normal, cauchy, inv_gamma, bernoulli_logit, gamma, laplace_marginal_*_lpmf %
	},
	emphstyle=\color{codepurple},%
	basicstyle={\ttfamily},
	breaklines=true,                 
	keepspaces=true,                 
	showspaces=false,                
}
\title{Hamiltonian Monte Carlo using an adjoint- differentiated Laplace approximation:
Bayesian inference for latent Gaussian models and beyond}

\author{%
  Charles C.~Margossian \\
  Department of Statistics \\
  Columbia University \\
  New York, NY 10027 \\
  \texttt{charles.margossian@columbia.edu}
  \And
  Aki~Vehtari \\
  Department of Computer Science \\
  Aalto University \\
  02150 Espoo, Finland \\
  Finnish Center for Artificial Intelligence
  \And
  Daniel~Simpson \\
  Department of Statistical Sciences \\
  University of Toronto \\
  ON M5S, Canada \\
  \And
  Raj~Agrawal \\
  CSAIL \\
  Massachusetts Institute of Technology \\
  Cambridge, MA 02139
}

\begin{document}

\maketitle

\begin{abstract}
  Gaussian latent variable models are a key class of Bayesian hierarchical models
  with applications in many fields.
  Performing Bayesian inference on such models can be challenging as 
  Markov chain Monte Carlo algorithms struggle with the geometry of the resulting posterior
  distribution and can be prohibitively slow.
  An alternative is to use a
  Laplace approximation to marginalize out the latent Gaussian variables and then 
  integrate out the remaining hyperparameters using dynamic Hamiltonian Monte Carlo,
  a gradient-based Markov chain Monte Carlo sampler.
  To implement this scheme efficiently,
  we derive a novel adjoint method that propagates the minimal information needed
  to construct the gradient of the approximate marginal likelihood.
  This strategy yields a scalable differentiation method that is orders of magnitude faster
  than state of the art differentiation techniques when the hyperparameters are high dimensional.
  We prototype the method in the probabilistic programming framework Stan
  and test the utility of the embedded Laplace approximation on several models,
  including one where the dimension of the hyperparameter is $\sim$6,000.
  Depending on the cases, the benefits can include an alleviation of the geometric pathologies
  that frustrate Hamiltonian Monte Carlo and a dramatic speed-up.
\end{abstract}

\section{Introduction}

  Latent Gaussian models observe the following hierarchical structure:
$$
    \phi  \sim  \pi(\phi), \qquad 
    \theta \sim \Normal(0, K(\phi)), \qquad
    y  \sim  \pi(y \mid \theta, \phi).
$$
  Typically,  single observations $y_i$ are independently distributed 
  and only depend on a linear combination of the latent variables, 
  that is $\pi(y_i \mid \theta, \phi) = \pi (y_i \mid a_i^T \theta, \phi)$, 
  for some appropriately defined vectors $a_i$.
  This general framework finds a broad array of applications: Gaussian processes,
  spatial models, and multilevel regression models to name a few examples.
  We denote $\theta$ the \textit{latent Gaussian variable} and $\phi$ the \textit{hyperparameter},
  although we note that in general $\phi$ may refer to any latent variable other than $\theta$.
  Note that there is no clear consensus in the literature on what constitutes a ``latent Gaussian model'';
  we use the definition from the seminal work by \citet{Rue:2009}.
  
  We derive a method to perform Bayesian inference on latent Gaussian models,
  which scales when $\phi$ is high dimensional and can handle the case where $\pi(\phi \mid y)$ is multimodal,
  provided the energy barrier between the modes is not too strong.
  This scenario arises in, for example, general linear models with a regularized horseshoe prior \cite{Carvalho+Polson+Scott:2010:HS}
  and in sparse kernel interaction models \cite{Agrawal:2019}.
  The main application for these models is studies with a low number of observations but a high-dimensional covariate,
  as seen in genomics.
  
\begin{wrapfigure}[15]{r}{7cm}
    \center
    \includegraphics[width=5.5cm]{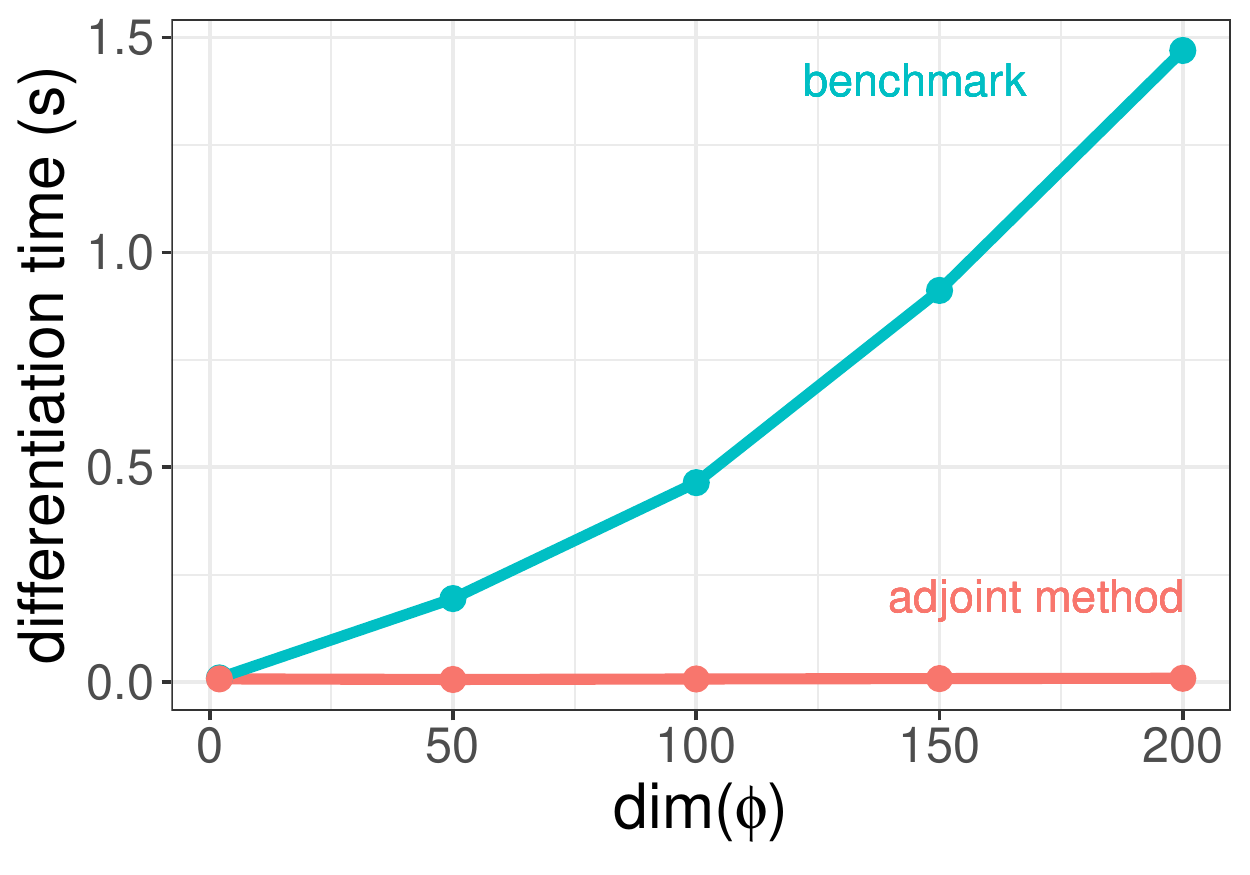}
    \vspace{-0.5cm}
    \caption{Wall time to differentiate the marginal density using the adjoint method
    (Algorithm~\ref{algo:diff2}) and, as a  benchmark,
    the method by \protect \citet{Rasmussen:2006} (Algorithm~\ref{algo:diff}).}
      \label{fig:compDiff}
  \end{wrapfigure}

  The inference method we develop uses a gradient-based Markov chain Monte Carlo (MCMC) sampler,
  coupled with a Laplace approximation to marginalize out $\theta$.
  The key to successfully implementing this scheme is a novel adjoint method
  that efficiently differentiates the approximate marginal likelihood. 
  In the case of a classic Gaussian process (Section~\ref{sec:gp}),
  where $\mathrm{dim}(\phi) = 2$,
  the computation required to evaluate and differentiate the marginal is on par
  with the GPstuff package \cite{Vanhatalo:2013},
  which uses the popular algorithm by \citet{Rasmussen:2006}.
  The adjoint method is however orders of magnitude faster when $\phi$ is high dimensional.
  Figure~\ref{fig:compDiff} shows the superior scalability of the adjoint method
  on simulated data from a sparse kernel interaction model.
  We lay out the details of the algorithms and the experiment in Section~\ref{sec:implementation}.
  
 \subsection{Existing methods}
  
  Bayesian computation  is, broadly speaking, split between two approaches:
  (i)  MCMC methods that approximately sample from the posterior,
  and (ii) approximation methods in which one finds a tractable distribution that approximates
  the posterior (e.g. variational inference, expectation propagation, and asymptotic approximations).
  The same holds for latent Gaussian models, where we can consider
  (i) Hamiltonian Monte Carlo (HMC) sampling \cite{Neal:2012, Betancourt:2018}
  and (ii) approximation schemes such as variational inference (VI) \cite{Blei:2017}
  or marginalizing out the latent Gaussian variables with a Laplace approximation
  before deterministically integrating the hyperparameters \cite{Tierney:1986, Rue:2009}.
  
  \paragraph{Hamiltonian Monte Carlo sampling.}
  When using MCMC sampling, the target distribution is
  \begin{equation*}
    \pi(\theta, \phi \mid y) \propto \pi(y \mid \theta, \phi) \pi(\theta \mid \phi) \pi(\phi),
  \end{equation*}
  and the Markov chain explores the joint parameter space of $\theta$ and $\phi$.

  HMC is a class of MCMC algorithms that powers many modern
  probabilistic programming languages, including Stan \cite{Carpenter:2017},
  PyMC3 \cite{Salvatier:2016}, and TensorFlow Probability~\cite{Dillon+etal:2017:tensorflow_distributions}.
  Its success is both empirically and theoretically motivated (e.g. \cite{Betancourt:2017})
  and, amongst other things, lies in its ability to probe the geometry of the target
  distribution via the gradient.
  The algorithm is widely accessible through a combination of its dynamic variants 
  \cite{Hoffman:2014, Betancourt:2018}, which spare the users the cumbersome task 
  of manually setting the algorithm's tuning parameters, and  {automatic differentiation}, 
  which alleviates the burden of calculating gradients by hand
  (e.g. \cite{Margossian:2019, Baydin:2018, Griewank:2008}).
  There are known challenges when applying HMC to hierarchical models,
  because of the posterior distribution's problematic geometry \cite{Betancourt:2013}.
  In the case of latent Gaussian models, this geometric grief is often caused by the latent Gaussian
  variable, $\theta$, and its interaction with $\phi$.
  Certain samplers, such as Riemannian HMC \cite{Girolami:2019, Betancourt:2013b} 
  and semi-separable HMC \cite{Zhang:2014}, are designed to better handle difficult geometries.
  While promising, these methods are difficult to implement, computationally expensive,
  and to our knowledge not widely used.

  \paragraph{Variational inference.} VI proposes to approximate the target distribution, $\pi(\theta, \phi \mid y)$,
  with a tractable distribution, $q(\theta, \phi)$, which minimizes the Kullback-Leibler divergence
  between the approximation and the target. 
  The optimization is performed over a pre-defined family of distributions, $\mathcal Q$.
  Adaptive versions, such as black-box VI \cite{Ranganath:2014}
  and automatic differentiation VI (ADVI) \cite{Kucukelbir:2017}, make it easy to run the algorithm.
  VI is further made accessible by popular software libraries,
  including the above-mentioned probabilistic programming languages,
  and others packages such as GPyTorch for Gaussian processes \cite{Gardner:2018}.
  For certain problems, VI is more scalable than MCMC, because 
  it can be computationally much faster to solve an optimization problem than to generate a large number of samples.
  There are however known limitations with VI (e.g. \cite{Blei:2017, Yao:2018, Huggins:2020, Dhaka:2020}).
  Of interest here is that $\mathcal Q$ may not include appropriate approximations of the target:
  mean field or full rank Gaussian families, for instance, will underestimate variance and settle on a single mode,
  even if the posterior is multimodal (e.g. \cite{Yao:2018}).

   \paragraph{Marginalization using a Laplace approximation.}
  The embedded Laplace approximation is a popular algorithm,
  and a key component of the R packages INLA
  (\textit{integrated nested Laplace integration}, \cite{Rue:2009,Rue:2017})
  and TMB (\textit{template model builder}, \cite{Kristensen:2016}),
  and the GPstuff package \cite{Vanhatalo:2013}.  
  The idea is to marginalize out $\theta$ and then use standard inference techniques on $\phi$.

    We perform the Laplace approximation
  \begin{eqnarray*}
    \pi(\theta \mid \phi, y) \approx \pi_\mathcal{G}(\theta \mid y, \phi) := \Normal(\theta^*, \Sigma^*),
  \end{eqnarray*}
  where $\theta^*$ matches the mode and $[\Sigma^*]^{-1}$ the curvature of $\pi(\theta \mid \phi, y)$.
  Then, the marginal posterior distribution is approximated as follows:
  \begin{eqnarray*}
    \pi(\phi \mid y) \approx \pi_\mathcal{G}(\phi \mid y) := \pi(\phi) \frac{\pi(\theta^* \mid \phi) \pi(y \mid \theta^*, \phi)}{\pi_\mathcal{G}(\theta^* \mid \phi, y) \pi(y)}.
  \end{eqnarray*}
  Once we perform inference on $\phi$, we can recover $\theta$ using the conditional
  distribution $\pi_\mathcal{G}(\theta \mid \phi, y)$ and effectively marginalizing $\phi$ out.
  For certain models, this yields much faster inference than MCMC,
  while retaining comparable accuracy \cite{Rue:2009}.
  Furthermore the Laplace approximation as a marginalization scheme 
  enjoys very good theoretical properties \cite{Tierney:1986}. 
  
  In the R package INLA, approximate inference is performed on $\phi$,
  by characterizing $\pi(\phi \mid y)$ around its presumed mode.
  This works well for many cases but presents two limitations:
  the posterior must be well characterized in the neighborhood of the estimated mode
  and it must be low dimensional, \textit{``2--5, not more than 20''} \cite{Rue:2017}.
  In one of the examples we study, the posterior of $\phi$ is both high dimensional
  ($\sim$6000) and multimodal.

  \paragraph{Hybrid methods.}  Naturally we can use a more flexible inference method on $\phi$
  such as a standard MCMC, as discussed by \citet{Gomez:2018},
  and  HMC as proposed in GPstuff and TMB,
  the latter through its extension  TMBStan and AdNuts 
  (\textit{automatic differentiation with a No-U-Turn Sampler} \cite{Monnahan:2018}).
  The target distribution of the HMC sampler is now $\pi_\mathcal{G}(\phi \mid y)$.

  To use HMC, we require the gradient of \mbox{$\log \pi_\mathcal{G}(y \mid \phi)$} with respect to $\phi$.
  Much care must be taken to ensure an efficient computation of this gradient.
  %
  TMB and GPstuff exemplify two approaches to differentiate the approximate marginal density.
  The first uses automatic differentiation and the second adapts the algorithms in 
  \citet{Rasmussen:2006}.
  One of the main bottlenecks is differentiating the estimated mode, $\theta^*$.
  %
  %
  In theory, it is straightforward to apply automatic differentiation,
  by brute-force propagating derivatives through $\theta^*$,
  that is, sequentially differentiating the iterations of a numerical optimizer.
  But this approach, termed the \textit{direct method}, is prohibitively expensive.
  A much faster alternative is to use the implicit function theorem
  (e.g. \cite{Bell:2008, Margossian:2019}).
  Given any accurate numerical solver,
  we can always use the implicit function theorem to get derivatives,
  as notably done in the Stan Math Library~\cite{Carpenter:2015} and in TMB's
  \textit{inverse subset algorithm} \cite{Kristensen:2016}.
  One side effect is that the numerical optimizer is treated as a black box.
  By contrast, \citet{Rasmussen:2006} define a bespoke Newton method to compute $\theta^*$,
  meaning we can store relevant variables from the final Newton step when computing derivatives.
  In our experience, this leads to important computational savings.
  But overall this method is much less flexible, 
  working well only when $\phi$ is low dimensional
  and requiring the user to pass the tensor of derivatives, $\partial K / \partial \phi$.
  %
  %

  \section{Aim and results of the paper}

  We improve the computation of HMC with an embedded Laplace approximation.
  Our implementation accommodates any covariance matrix $K$,
  without requiring the user to specify $\partial K / \partial \phi$,
  efficiently differentiates $\log \pi_\mathcal{G}(y \mid \phi)$, 
  even when $\phi$ is high dimensional,
  and deploys dynamic HMC to perform inference on $\phi$.
  We introduce a novel adjoint method to differentiate $\log \pi_\mathcal{G}(y \mid \phi)$,
  build the algorithm in C++, and add it to the Stan language.
  Our approach combines the  Newton solver of \citet{Rasmussen:2006}
  with a non-trivial application of automatic differentiation.
  
  Equipped with this implementation, we test dynamic HMC with an embedded Laplace
  approximation on a range of models,
  including ones with a high dimensional and multimodal hyperparameter.
  We do so by benchmarking our implementation against Stan's dynamic HMC,
  which runs MCMC on both the hyperparameter and the latent Gaussian variable.
  For the rest of the paper, we call this standard use of dynamic HMC, \textit{full HMC}.
  We refer to marginalizing out $\theta$ and using dynamic HMC on $\phi$, 
  as the \textit{embedded Laplace approximation}.
  Another candidate benchmark is Stan's ADVI.
  \citet{Yao:2018} however report that ADVI underestimates the posterior variance
  and returns a unimodal approximation, even when the posterior is multimodal.
  We observe a similar behavior in the models we examine.
  For clarity, we relegate most of our analysis on ADVI to the Supplementary Material.
  
  Our computer experiments identify cases where the benefits of the embedded Laplace approximation, 
  as tested with our implementation, are substantial.
  In the case of a classic Gaussian process,
  with $\mathrm{dim}(\phi) = 2$ and $\mathrm{dim}(\theta) = 100$,
  we observe an important computational speed up, when compared to full HMC.
  We next study a general linear regression with a sparsity inducing prior;
  this time $\mathrm{dim}(\phi) \approx 6,000$ and $\mathrm{dim}(\theta) \approx 100$.
  Full HMC struggles with the posterior's geometry, as indicated by divergent transitions,
  and requires a model reparameterization and extensive tuning of the sampler.
  On the other hand, the embedded Laplace approximation evades many of the geometric
  problems and solves the approximate problem efficiently.
  We observe similar results for a sparse kernel interaction model,
  which looks at second-order interactions between covariates \cite{Agrawal:2019}.
  Our results stand in contrast to the experiments presented in \citet{Monnahan:2018}, 
  who used a different method to automatically differentiate the Laplace approximation 
  and reported at best a minor speed up.
  We do however note that the authors investigated different models than the ones we study here.
  
  In all the studied cases, the likelihood is log-concave.
  Combined with a Gaussian prior, log-concavity guarantees that $\pi(\theta \mid \phi, y)$ is unimodal.
  Detailed analysis on the error introduced by the Laplace approximation
  for log-concave likelihoods
  can be found in references 
  (e.g. \cite{Kuss:2005, Vanhatalo+Pietilainen+Vehtari:2010, Cseke:2011, Vehtari+etal:2016:loo_glvm})
  and are consistent with the results from our computer experiments.

  \section{Implementation for probabilistic programming}
  \label{sec:implementation}
  
  In order to run HMC, we need a function that returns the approximate 
  log density of the marginal
  likelihood, $\log \pi_\mathcal{G}(y \mid \phi)$, and its gradient with respect to $\phi$, 
  $\nabla_\phi \log \pi_\mathcal{G}(y \mid \phi)$.
  The user specifies the observations, $y$, and
  a function to generate the covariance $K$, based on input covariates $x$
  and the hyperparameters $\phi$.
  In the current prototype, the user picks the likelihood, $\pi(y \mid \theta, \phi)$, from
  a set of options\footnote{
    More likelihoods can be implemented through a C++ class that specifies the first three derivatives of the log-likelihood.}:
  for example, a likelihood arising from a Bernoulli distribution with a logit link.
  
  Standard implementations of the Laplace approximation use the algorithms in 
  \citet[chapter~3 and 5]{Rasmussen:2006} 
  to compute (i) the mode $\theta^*$ and $\log \pi_\mathcal{G}(y \mid \phi)$,
  using a Newton solver;
  (ii) the gradient $\nabla_\phi \log \pi_\mathcal{G}(y \mid \phi)$ (Algorithm~\ref{algo:diff}),
  and (iii) simulations from $\pi_\mathcal{G}(\theta \mid y, \phi)$.
 The major contribution of this paper is to construct a new differentiation algorithm, i.e. item (ii).

   \subsection{Using automatic differentiation in the algorithm of \citet{Rasmussen:2006}}

   The main difficulty with Algorithm~\ref{algo:diff} from \citet{Rasmussen:2006} 
   is the requirement for $\partial K / \partial \phi_j$
   %
   at line 8. For classic problems, where $K$ is, for instance, a exponentiated quadratic kernel,
   the derivatives are available analytically.
   This is  not the case in general
   and, in line with the paradigm of probabilistic programming,
   we want a method that does not require the user to specify 
   the tensor of derivatives, $\partial K / \partial \phi$.
  
    \begin{algorithm}
    \caption{\textit{Gradient of the approximate marginal density, $\pi_\mathcal{G}(y \mid \phi)$,
    with respect to the 
    hyperparameters $\phi$}, adapted from algorithm 5.1 by \protect
    \citet[chapter~5]{Rasmussen:2006}.
    We store and reuse terms computed during the final Newton step,
    algorithm 3.1 in \protect \citet[chapter~3]{Rasmussen:2006}.}
    \label{algo:diff}
    \begin{algorithmic}[2]
    \item \textbf{input:} $y$, $\phi$, $\pi(y \mid \theta, \phi)$
    \item \textbf{saved input from the  Newton solver}: $\theta^*$, $K$, $W^\frac{1}{2}$, $L$, $a$
    \item $Z = \frac{1}{2} a^T \theta^* + \log \pi(y \mid \theta^*, \phi) - \sum \log (\mathrm{diag}(L))$
    \item $R = W^\frac{1}{2} L^T \setminus (L \setminus W^\frac{1}{2})$
    \item $C = L \setminus (W^\frac{1}{2} K)$
    \item $s_2 = - \frac{1}{2} \mathrm{diag} (\mathrm{diag}(K) - \mathrm{diag}(C^T C))
      \nabla_\theta^3 \log \pi(y \mid \theta^*, \phi)$
    \item \textbf{for} $j = 1$ ... $\mathrm{dim}(\phi)$
    \item \hspace*{0.5cm} $K' = \partial K / \partial \phi_j$
    \item \hspace*{0.5cm} $s_1 = \frac{1}{2} a^T K' a - \frac{1}{2} \mathrm{tr}(RK')$
    \item \hspace*{0.5cm} $b = K' \nabla_\theta \log \pi(y \mid \theta, \phi)$
    \item \hspace*{0.5cm} $s_3 = b - K R b$
    \item \hspace*{0.5cm} $\frac{\partial}{\partial \phi_j} \pi(y \mid \phi) = s_1 + s_2^T s_3$
    \item \textbf{end for}
    \item \textbf{return} $\nabla_\phi \log \pi_\mathcal{G}(y \mid \phi)$ 
    \end{algorithmic}
  \end{algorithm}
  
   Automatic differentiation allows us to numerically evaluate $\partial K / \partial \phi$
   based on computer code to evaluate $K$.
   To do this, we introduce the map $\mathcal K$
   \begin{eqnarray*}
     \mathcal K & : & \mathbb R^p \to \mathbb R^{n (n + 1) / 2} \\
       & & \phi \to K,
   \end{eqnarray*}
   where $p$ is the dimension of $\phi$ and $n$ that of $\theta$.
   To obtain the full tensor of derivatives, we require either $p$ forward mode sweeps
   or $n (n + 1) / 2$ reverse mode sweeps.
   Given the scaling, we favor forward mode and this works well when $p$ is small.
   However, once $p$ becomes large, this approach is spectacularly inefficient.
  
   \subsection{Adjoint method to differentiate the approximate log marginal density}
   \label{sec:adjoint}
   
   To evaluate the gradient of a composite map,
   it is actually not necessary to compute the full Jacobian matrix of intermediate operations.
   This is an important, if often overlooked, property of automatic differentiation 
   and the driving principle behind \textit{adjoint methods} (e.g. \cite{Errico:1997}).
   This idea motivates an algorithm that does not explicitly construct $\partial K / \partial \phi$,
   a calculation that is both expensive and superfluous.
   Indeed, it suffices to evaluate $w^T \partial K / \partial \phi$ for the correct \textit{cotangent vector}, $w^T$,
   an operation we can do in a single reverse mode sweep of automatic differentiation.

  \begin{theorem}
  \label{thm:adjoint}
  Let $ \log \pi_\mathcal{G}(y \mid \phi)$ be the approximate log marginal density
  in the context of a latent Gaussian model.
  Let $a$ be defined as in the  Newton solver by \citet[chapter~3]{Rasmussen:2006},
  and let $R$ and $s_2$ be defined as in Algorithm~\ref{algo:diff}.
  Then
  \begin{equation*}
    \nabla_\phi \log \pi_\mathcal{G}(y \mid \phi) = w^T \frac{\partial K}{\partial \phi},
  \end{equation*} 
  where the gradient is with respect to $\phi$ and
  \begin{equation*}
    w^T = \frac{1}{2} aa^T - \frac{1}{2} R + (s_2 + RK s_2) [\nabla_\theta \log \pi(y \mid \theta, \phi)]^T.
  \end{equation*}
  \end{theorem}
  
  The proof follows from Algorithm~\ref{algo:diff} and noting that all the operations in 
  $\partial K / \partial \phi_j$ are linear.
  We provide the details in the Supplementary Material.
  Armed with this result, we build Algorithm~\ref{algo:diff2},
  a method that combines the insights of \citet{Rasmussen:2006} with the principles of adjoint methods.
   
    \begin{algorithm}
    \caption{\textit{Gradient of the approximate marginal log density, $\log \pi_\mathcal{G}(y \mid \phi)$, 
    with respect
    to the hyperparameters, $\phi$,} using reverse mode automatic differentiation
    and theorem~\ref{thm:adjoint}.}
    \label{algo:diff2}
    \begin{algorithmic}[2]
    \item \textbf{input:} $y$, $\phi$, $\pi(y \mid \theta, \phi)$
    \item Do lines 2 - 6 of Algorithm~\protect \ref{algo:diff}.
    \item Initiate an expression graph for automatic differentiation with $\phi_v = \phi$.
    \item $K_v = \mathcal K(\phi_v)$
    \item
    $w^T = \frac{1}{2} a a^T - \frac{1}{2} R + (s_2 + R K s_2) [\nabla_\theta \log \pi(y \mid \theta, \phi)]^T$
    \item Do a reverse sweep over $K$, with $w^T$ as the initial cotangent to obtain
    $\nabla_\phi \log \pi_\mathcal{G}(y \mid \phi)$.
    \item \textbf{return:} $\nabla_\phi \log \pi_\mathcal{G}(y \mid \phi)$.
    \end{algorithmic}
  \end{algorithm}
  

  Figure \ref{fig:compDiff} shows the time required for one evaluation and differentiation of $\log \pi_\mathcal{G}(y \mid \phi)$
  for the sparse kernel interaction model developed by \citet{Agrawal:2019}
  on simulated data.
 The covariance structure of this model is nontrivial and analytical derivatives are not easily  available.
  We simulate a range of data sets for varying dimensions, $p$, of $\phi$.
  For low dimensions, the difference is small; however, for $p = 200$,
  Algorithm~\ref{algo:diff2} is more than 100 times faster than Algorithm~\ref{algo:diff},
  requiring 0.009 s, instead of 1.47 s.

   \section{Gaussian process with a Poisson likelihood}
   \label{sec:gp}
   
   We fit the disease map of Finland by \citet{Vanhatalo+Pietilainen+Vehtari:2010} which models the mortality
   count across the country.
   The data is aggregated in $n = 911$ grid cells.
   We use 100 cells, which allows us to fit the model quickly
   both with full HMC and HMC using an embedded Laplace approximation.
   For the $i^\mathrm{th}$ region, we have a 2-dimensional coordinate $x_i$,
   the counts of deaths $y_i$, and the standardized expected number of deaths, $y_e^i$.
   %
    %
    The full latent Gaussian model is
 \begin{eqnarray*}
      (\rho, \alpha)  \sim  \pi(\rho, \alpha), \qquad
      \theta  \sim  \Normal(0, K(\alpha, \rho, x)), \qquad
      y_i  \sim  \mathrm{Poisson}(y_e^i e^{\theta_i}),
\end{eqnarray*}
  where $K$ is an exponentiated quadratic kernel, $\alpha$ is the marginal standard deviation
  and $\rho$ the characteristic length scale. Hence $\phi = (\alpha, \rho)$.

  Fitting this model with MCMC requires running the Markov chains over $\alpha$, $\rho$, and $\theta$.
  Because the data is sparse --- one observation per group ---
  the posterior has a funnel shape which can
  lead to biased MCMC estimates \cite{Neal:2003, Betancourt:2013}. 
  A  useful diagnostic for identifying posterior shapes that challenge the HMC sampler 
  is \textit{divergent transitions}, 
  which occur when there is significant numerical error in the computation of
  the Markov chain trajectory \cite{Betancourt:2018}.
  
  To remedy these issues, we reparameterize the model and adjust the \textit{target acceptance rate}, 
  $\delta_a$.
  $\delta_a$ controls the precision of HMC, with the usual trade-off between accuracy and speed.
  For well behaved problems, the optimal value is 0.8 \cite{Betancourt:2015} but posteriors with highly varying curvature require a higher value.
  Moreover, multiple attempts at fitting the model must be done before we correctly tune the sampler
  and remove all the divergent transitions.
  See the Supplementary Material for more details.
  
  An immediate benefit of the embedded Laplace approximation is that we marginalize out $\theta$
  and only run HMC on $\alpha$ and $\rho$, a two-dimensional and typically
  well behaved parameter space.
  In the case of the disease map, we do not need to reparameterize the model, nor adjust $\delta_a$.
  
\begin{figure}[tbp]
    \begin{center}
      \includegraphics[width=5.5in]{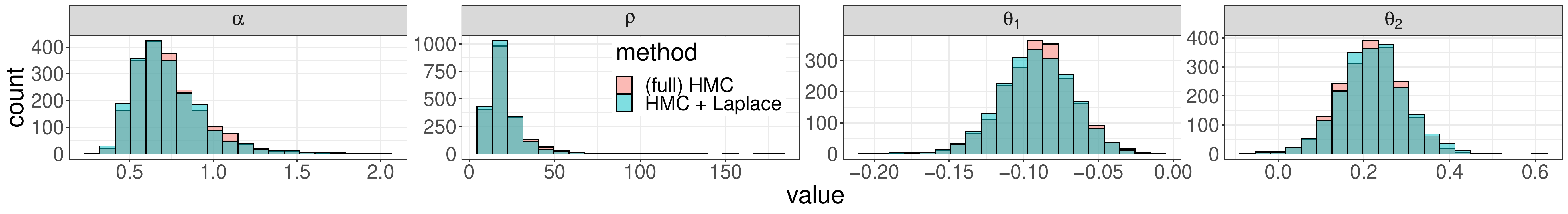}
      \\ \ \\
      \includegraphics[width=5.5in]{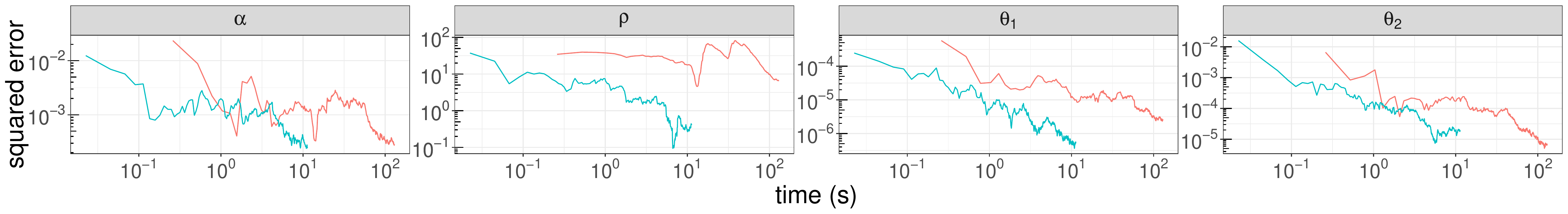}
    \end{center}
    \caption{(Up) Posterior samples obtained with full HMC and the embedded Laplace approximation
    when fitting the disease map.
    (Down) Error when estimating the expectation value against wall time.
    Unreported in the figure is that we had to fit full HMC twice before obtaining good tuning parameters.}
    \label{fig:gp_comp}
  \end{figure}

  We fit the models with both methods, using 4 chains, each with 500 warmup and 500 sampling iterations.
  A look at the marginal distributions of $\alpha$, $\rho$, and the first two elements of $\theta$
  suggests the posterior samples generated by full HMC and the embedded Laplace approximation
  are in close agreement (Figure~\ref{fig:gp_comp}).
  With a Poisson likelihood, the bias introduced by the Laplace approximation is small,
  as  shown by \citet{Vanhatalo+Pietilainen+Vehtari:2010}.
  We benchmark the Monte Carlo estimates of both methods against results from running 18,000
  MCMC iterations.
  The embedded Laplace approximations yields comparable precision, when estimating expectation values,
  and is an order of magnitude faster (Figure~\ref{fig:gp_comp}).
  In addition, we do not need to tune the algorithm and the MCMC warmup time is much shorter
  ($\sim$10 seconds against $\sim$200 seconds for full HMC).
  
  \section{General linear regression model with a regularized horseshoe prior}

  Consider a regression model with $n$ observations and $p$ covariates.
  In the ``$p \gg n$'' regime, we typically need additional structure, such as sparsity,
  for accurate inference. 
  The {horseshoe prior} \cite{Carvalho+Polson+Scott:2010:HS} is a useful prior
  when it is assumed that only a small portion of the regression coefficients 
  are  non-zero.
  Here we use the regularized horseshoe prior by \citet{Piironen:2017}.
  The horseshoe prior is parameterized by a global scale term, the scalar $\tau$,
  and local scale terms for each covariate, $\lambda_j$, $j=1,\ldots, p$.
  Consequently the number of hyperparameters is $\mathcal O(p)$.
  
  To use the embedded Laplace approximation, 
  we recast the regularized linear regression as a latent Gaussian model.
  The benefit of the approximation is not a significant speedup,
  rather an improved posterior geometry, due to marginalizing $\theta$ out.
  This means we do not need to reparameterize the model, nor fine tune the sampler.
   To see this, we examine the genetic microarray classification data set on prostate cancer
   used by \citet{Piironen:2017} and fit a regression model with a Bernoulli distribution and a logit link.
   Here, $\mathrm{dim}(\theta) = 102$ and $\mathrm{dim}(\phi) = 5,966$.

  We use 1,000 iterations to warm up the sampler and 12,000 sampling iterations.
  Tail quantiles, such as the $90^\mathrm{th}$ quantile,
  allow us to identify parameters which have a small local shrinkage and thence indicate relevant covariates.
  The large sample size is used to reduce the Monte Carlo error in our estimates of these extreme quantiles.

   \begin{wrapfigure}[21]{r}{2.3in}
     \center
    \includegraphics[width=2.25in]{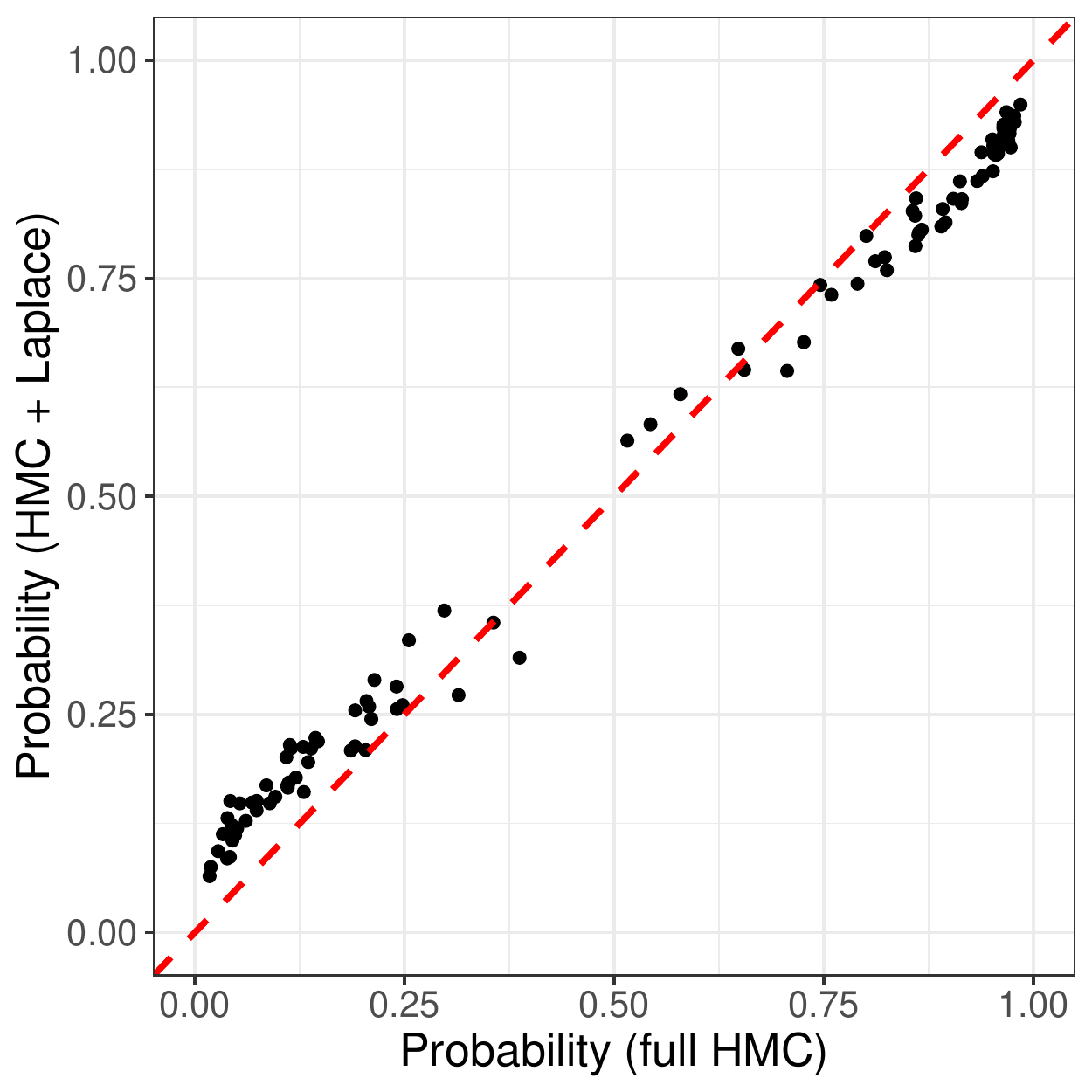}
    \vspace{-0.2cm}
    \caption{Expectation value for the probability of developing prostate cancer,
    as estimated by full HMC and HMC using an embedded
    Laplace approximation.}
    \label{fig:hs_prob_comp}
 \end{wrapfigure} 

  Fitting this model with full HMC requires a fair amount of work:
  the model must be reparameterized and the sampler carefully tuned,
  after multiple attempts at a fit.
  We use a non-centered parameterization, set $\delta_a = 0.999$
  (after attempting $\delta_a = 0.8$ and $\delta_a = 0.99$) and do some additional adjustments.
  Even then we obtain 13 divergent transitions over 12,000 sampling iterations.
  The Supplementary Material describes the tuning process in all its thorny details.
  By contrast, running the embedded Laplace approximation with Stan's default tuning parameters
  produces 0 divergent transitions.
  Hence the approximate problem is efficiently solved by dynamic HMC.
  Running ADVI on this model is also straightforward.
  
\begin{figure}[tbp]
    \begin{center}
      \includegraphics[width=5.5in]{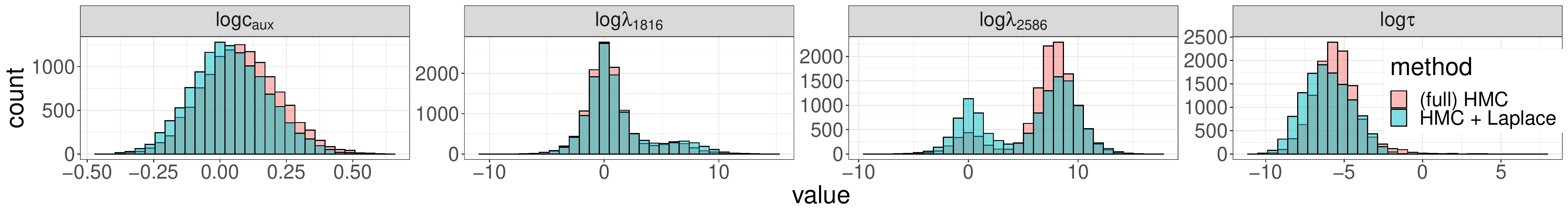}
      \includegraphics[width=5.5in]{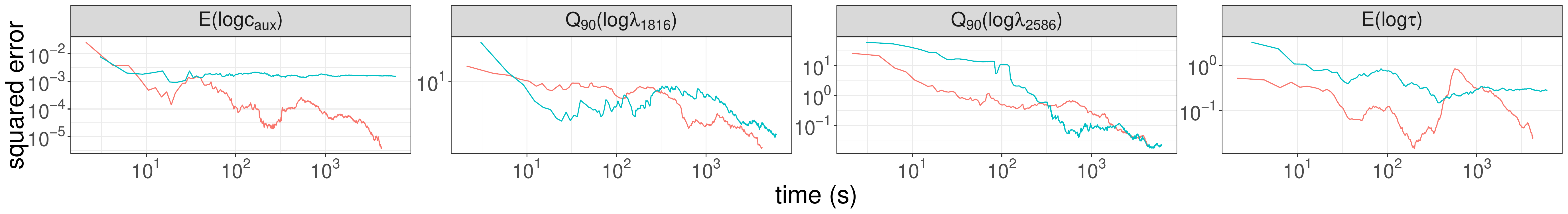}
    \end{center}
    \caption{(Up) Posterior samples obtained with full HMC and HMC using an embedded Laplace approximation
    when fitting a general linear regression with a regularized horseshoe prior.
    (Down) Error when estimating various quantities of interest against wall time.
    $E$ stands for ``expectation'' and $Q_{90}$, ``90$^\mathrm{th}$ quantile''.
    Unreported in the figure is that we had to run full HMC four times before obtaining reasonable tuning parameters.}
    \label{fig:horseshoe_comp}
  \end{figure}

  Table~\ref{tab:lambda} shows the covariates with the highest $90^\mathrm{th}$ quantiles,
  which are softly selected by full HMC, the embedded Laplace approximation and ADVI.
  For clarity, we exclude ADVI from the remaining figures but note that it generates, for this particular problem,
  strongly biased inference; more details can be found in the Supplementary Material.
  Figure~\ref{fig:hs_prob_comp} compares the expected probability of developing cancer.
  Figure~\ref{fig:horseshoe_comp} compares the posterior samples
  and the error when estimating various quantities of interest,
  namely (i) the expectation value of the global shrinkage, $\tau$,
  and the slab parameter, $c_\mathrm{aux}$;
  and (ii) the $90^\mathrm{th}$ quantile of two local shrinkage parameters.
  As a benchmark we use estimates obtained from 98,000 MCMC iterations.

  \begin{table}
    \renewcommand{\arraystretch}{1.5}
    \begin{center}
    \caption{Top six covariate indices, $i$, 
    with the highest $90^\mathrm{th}$ quantiles of $\log \lambda_i$
    for the general linear model with a regularized horseshoe prior.
    The first two methods are in good agreement; ADVI selects different covariates,
    in part because it approximates the multimodal posterior with a unimodal distribution (see the Supplementary Material).}
    \label{tab:lambda}
    \begin{tabular} {l r r r r r r}
    \rowcolor{pink}
    \textbf{(full) HMC} & 2586 & 1816 & 4960 & 4238 & 4843 & 3381 \\
    \rowcolor{LightCyan}
    \textbf{HMC + Laplace} & 2586 & 1816 & 4960 & 4647 & 4238 & 3381 \\
    \textbf{ADVI} & 1816 & 2416 & 4284 & 2586 & 5279 & 4940
    \end{tabular}
    \end{center}
  \end{table}
  
  The Laplace approximation 
  yields slightly less extreme probabilities of developing cancer than the corresponding full model.
  This behavior is expected for latent Gaussian models with a Bernoulli observation model,
  and has been studied in the cases of  Gaussian processes
  and Gaussian random Markov fields
  (e.g. \cite{Kuss:2005, Cseke:2011, Vehtari+etal:2016:loo_glvm}).
  While introducing a bias,
  the embedded Laplace approximation yields accuracy comparable to full HMC
  when evaluating quantities of interest.

  \section{Sparse kernel interaction model}
  
  A natural extension of the general linear model is to include interaction terms.
  To achieve better computational scalability, we can use the {kernel interaction trick}
  by \citet{Agrawal:2019} and build a {sparse kernel interaction model} (SKIM),
  which also uses the regularized horseshoe prior by \citet{Piironen:2017}.
  The model is an explicit latent Gaussian model  and uses a non-trivial covariance matrix.
  The full details of the model are given in the Supplementary Material.
  
  
  When fitting the SKIM to the prostate cancer data, 
  we encounter similar challenges as in the previous section:
  $\sim$150 divergent transitions with full HMC when using Stan's default tuning parameters.
  The behavior when adding the embedded Laplace approximation is much better,
  although there are still $\sim$3 divergent transitions,\footnote{
  We do our preliminary runs using only 4000 sampling iterations.
  The above number are estimated for 12000 sampling iterations.
  The same holds for the estimated run times.} 
  which indicates that this problem remains quite difficult even after the approximate marginalization.
  We also find large differences in running time.
  The embedded Laplace approximation runs for $\sim$10 hours,
  while full HMC takes $\sim$20 hours with $\delta_a = 0.8$
  and $\sim$50 hours with $\delta_a = 0.99$, making it difficult to tune the sampler
  and run our computer experiment.
  
    \begin{table}
    \renewcommand{\arraystretch}{1.5}
    \begin{center}
    \caption{Top six covariate indices, $i$, 
    with the highest $90^\mathrm{th}$ quantiles of $\log \lambda_i$
    for the SKIM.}
    \label{tab:skim}
    \begin{tabular} {l r r r r r r}
    \rowcolor{pink}
    \textbf{(full) HMC} & 2586 & 2660 & 2679 & 2581 & 2620 & 2651 \\
    \rowcolor{LightCyan}
    \textbf{HMC + Laplace} & 2586 & 2679 & 2660 & 2581 & 2620 & 2548 \\
    \textbf{ADVI} & 2586 & 2526 & 2106 & 2550 & 2694 & 2166 \\
    \end{tabular}
    \end{center}
  \end{table}
 
  For computational convenience, we fit the SKIM
  using only 200 covariates, indexed 2500 - 2700 to encompass the 2586$^\mathrm{th}$
  covariate which we found to be strongly explanatory.
  This allows us to easily tune full HMC without altering the takeaways of the experiment.
  Note that the data here used is different from the data we used in the previous section
  (since we only examine a subset of the covariates)
  and the marginal posteriors should therefore not be compared directly.
  
\begin{figure}[tbp]
    \begin{center}
      \includegraphics[width=5.5in]{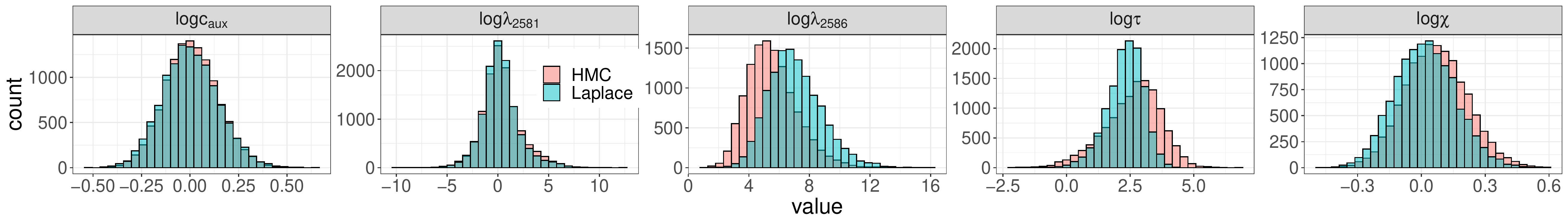}
      \includegraphics[width=5.5in]{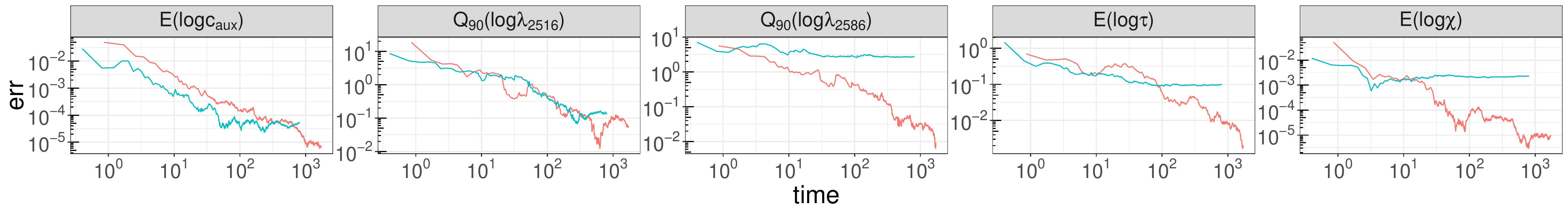}
    \end{center}
    \caption{(Up) Samples obtained with full HMC and HMC using an embedded Laplace approximation
    when fitting the SKIM.
    (Down) Error when estimating various quantities of interest against wall time.
    $E$ stands for ``expectation'' and $Q_{90}$, ``90$^\mathrm{th}$ quantile''.
    Unreported in the figure is that we had to run full HMC twice before obtaining reasonable tuning parameters.}
    \label{fig:skim_comp}
  \end{figure}

  As in the previous section, we generate 12,000 posterior draws for each method.
  For full HMC we obtain 36 divergent transitions with $\delta_a = 0.8$ and 0 with $\delta_a = 0.99$.
  The embedded Laplace approximation produces 0 divergences with $\delta_a = 0.8$.
  Table~\ref{tab:skim} shows the covariates which are softly selected.
  As before, we see a good overlap between full HMC and the embedded Laplace approximation,
  and mostly disagreeing results from ADVI.
  Figure~\ref{fig:skim_comp} compares (i)  the posterior draws of full HMC and the embedded Laplace approximation,
  and (ii) the error over time, benchmarked against estimates from 98,000 MCMC iterations,
  for certain quantities of interest.
  We obtain comparable estimates but note that the Laplace approximation introduces a bias,
  which becomes more evident over longer runtimes.
  
  \section{Discussion}
  
  Equipped with a scalable and flexible differentiation algorithm, we expand the regime of models to which
  we can apply the embedded Laplace approximation.
  HMC allows us to perform inference even when $\phi$ is high dimensional and multimodal,
  provided the energy barrier is not too strong.
  In the case where $\mathrm{dim}(\theta) \gg \mathrm{dim}(\phi)$,
  the approximation also yields a dramatic speedup.
  When $\mathrm{dim}(\theta) \ll \mathrm{dim}(\phi)$, marginalizing $\theta$ out
  can still improve the geometry of the posterior,
  saving the user time otherwise spent tuning the sampling algorithm.
  However, when the posterior is well-behaved, the approximation may not provide any benefit.
  
  Our next step is to further develop the prototype for Stan.
  We are also aiming to incorporate features that allow for
  a high performance implementation,
  as seen in the packages INLA, TMB, and GPstuff.
  Examples include support for sparse matrices required to fit latent Markov random fields,
  parallelization and GPU support.
  
  We also want to improve the flexibility of the method by allowing users
  to specify their own likelihood.
  TMB provides this flexibility but in our view two important challenges persist.
  Recall that unlike full HMC, which only requires first-order derivatives,
  the embedded Laplace approximation requires the third-order derivative of the likelihood
  (but not of the other components in the model).
  It is in principle possible to apply automatic differentiation to evaluate higher-order derivatives
  and most libraries, including Stan, support this;
  but, along with feasibility, there is a question of efficiency and practicality
  (e.g. \cite{Betancourt:2018b}):
  the automated evaluation of higher-order derivatives is often prohibitively expensive.
  The added flexibility also burdens us with more robustly diagnosing errors
  induced by the approximation.
  There is extensive literature on log-concave likelihoods
  but less so for general likelihoods.
  Future work will investigate diagnostics such as importance sampling \cite{Vehtari+etal:2019:psis}, 
  leave-one-out cross-validation 
  \cite{Vehtari+etal:2016:loo_glvm}, and simulation based calibration \cite{Talts:2018}.
  
  \section*{Broader Impact}
  
  Through its multidisciplinary nature, the here presented research can act
  as a bridge between various communities of statistics and machine learning.
  We hope practitioners of MCMC will consider the benefits
  of approximate distributions and vice-versa.
  This work may be a stepping stone to a broader conversation on how,
  what we have called the two broad approaches of Bayesian computation,
  can be combined.
  The paper also raises awareness about existing technologies
  and may dispel certain misconceptions.
  For example, our use of the adjoint principle shows
  that automatic differentiation is not a simple application of the chain rule,
  but quite a bit more clever than that.
    
  Our goal is to make the method readily available to practitioners
  across multiple fields,
  which is why our C++ code and prototype Stan interface are open-source.
  While there is literature on the Laplace approximation,
  the error it introduces, and the settings in which it works best,
  we realize not all potential users will be familiar with it.
  To limit misuse, we must complement our work with
  pedagogical material built on the existing references,
  as well as support and develop more diagnostic tools.  

\section*{Acknowledgment}
We thank Michael Betancourt, Steve Bronder, Alejandro Catalina, Rok \u{C}e\u{s}novar, Hyunji Moon, Sam Power, Sean Talts and Yuling Yao for helpful discussions.

CM thanks the Office of Naval Research, the National Science Foundation, the Institute for Education
Sciences, and the Sloan Foundation. CM and AV thank the Academy of Finland (grants 298742 and
313122). DS thanks the Canada Research Chairs program and the Natural Sciences and Engineering
Research Council of Canada. RA's research was supported in part by a grant from DARPA.

We acknowledge computing resources from Columbia University’s Shared Research Computing
Facility project, which is supported by NIH Research Facility Improvement Grant 1G20RR030893-
01, and associated funds from the New York State Empire State Development, Division of Science
Technology and Innovation (NYSTAR) Contract C090171, both awarded April 15, 2010.

We also acknowledge the computational resources provided by the Aalto Science-IT project.

The authors declare to have no conflict of interest.

\appendix

\section*{Appendix (Supplementary Material)}

We review the Newton solver proposed by \citet{Rasmussen:2006}
and prove theorem~\ref{thm:adjoint}, the main result required to do build an adjoint method
for the embedded Laplace approximation.
We next present our prototype code
and provide details for the models used in our computer experiments.

\section{Newton solver for the embedded Laplace approximation}

    \begin{algorithm}[!h]
    \caption{\textit{Newton solver for the embedded Laplace approximation}
    \protect\cite[chapter~3]{Rasmussen:2006}}
    \label{algo:mode}
      \begin{algorithmic}[2]
      \item \textbf{input:} $K$, $y$, $\pi(y \mid \theta, \phi)$
      \item $\theta^* = \theta_0$ \hspace{1cm} (initialization)
      \item \textbf{repeat}
      \item \hspace*{0.5cm} $W = - \nabla_\theta \nabla_\theta \log \pi(y \mid \theta^*, \phi)$
      \item \hspace*{0.5cm} $L = \mathrm{Cholesky}(I + W^\frac{1}{2} K W^\frac{1}{2})$
      \item \hspace*{0.5cm} $b = W \theta^* + \nabla_\theta \log \pi(y \mid \theta^*, \phi)$
      \item \hspace*{0.5cm} $a = b - W^\frac{1}{2} L^T \setminus (L \setminus (W^\frac{1}{2} K b))$
      \item \hspace*{0.5cm} $ \theta^* = Ka$
      \item \textbf{until} convergence
      \item $\log \pi(y \mid \phi) = -\frac{1}{2} a^T \theta^* + \log \pi(y \mid \theta^*, \phi) - \sum_i \log L_{ii}$
    \item \textbf{return:} $\theta^*$, $\log \pi_\mathcal{G}(y \mid \phi)$
    \end{algorithmic}
  \end{algorithm}

  Algorithm~\ref{algo:mode} is a transcription of the Newton method 
  by \citet[chapter~3]{Rasmussen:2006} using our notation.
  As a convergence criterion,
  we use the change in the objective function between two iterations
  $$
    \Delta \log \pi(\theta \mid y, \phi) \le \epsilon
  $$
  for a specified $\epsilon$. This is consistent with the approach
  used in GPStuff \cite{Vanhatalo:2013}.
  We store the following variables generated during the final Newton step
  to use them again when computing the gradient:
  $\theta^*$, $K$, $W^{\frac{1}{2}}$, $L$, and $a$.
  This avoids redundant computation and spares us an expensive
  Cholesky decomposition.
  
\section{Building the adjoint method}
  
   To compute the gradient of the approximate log marginal with respect to $\phi$,
   $\nabla \log \pi_\mathcal{G}(y \mid \phi)$,
   we exploit several important principles of automatic differentiation.
   While widely used in statistics and machine learning,
   these principles remain arcane to many practitioners and deserve a brief review.
   We will then construct the adjoint method (theorem~\ref{thm:adjoint} and algorithm~\ref{algo:diff2})
   as a correction to algorithm~\ref{algo:diff}.
   
   \subsection{Automatic differentiation}

   Given a composite map
  \begin{eqnarray*}
    f = f^L \circ f^{L - 1} \circ ... f^1,
  \end{eqnarray*}
  the chain rule teaches us that the corresponding Jacobian matrix observes a similar decomposition:
  \begin{eqnarray*}
    J = J_L  \cdot J_{L - 1} \cdot ... \cdot J_1.
  \end{eqnarray*}
  Based on computer code to calculate $f$,
  a \textit{forward mode sweep} automatic differentiation numerically
  evaluates the action of the Jacobian matrix on the initial tangent $u$,
  or \textit{directional derivative} $J \cdot u$.
  Extrapolating from the chain rule
  \begin{eqnarray*}
    \begin{aligned}
    J \cdot u & = J_L \cdot J_{L - 1} \cdot ... \cdot J_3 \cdot J_2 \cdot J_1 \cdot u \\
                   & = J_L \cdot J_{L - 1} \cdot ... \cdot J_3 \cdot J_2 \cdot u_1 \\
                   & = J_L \cdot J_{L - 1} \cdot ... \cdot J_3 \cdot u_2 \\
                   & . . . \\
                   & = J_L \cdot u_{L - 1}, \\
    \end{aligned}
  \end{eqnarray*}
  where the $u_l$'s verify the recursion relationship
  \begin{eqnarray*}
    \begin{aligned}
    u_1 & = J_1 \cdot u, \\
    u_l & = J_l \cdot u_{l - 1}. \\
    \end{aligned}
  \end{eqnarray*}
  If our computation follows the steps outlined above
  we never need to explicitly compute the full Jacobian matrix, $J_l$, 
  of an intermediate function, $f^l$;
  rather we only calculate a sequence of Jacobian-tangent products.
  Similarly a \textit{reverse mode sweep} evaluates the contraction of the Jacobian matrix 
  with a cotangent, $w^T$, yielding $w^T J$,
  by computing a sequence cotangent-Jacobian products.
  
  Hence, in the case of the embedded Laplace approximation,
  where 
  \begin{eqnarray*}
    \mathcal K : & \phi & \to K \\
                        & \mathbb R^p & \to \mathbb R^{(n + 1) n / 2}
  \end{eqnarray*}
  is an intermediate function,
  we do not need to explicitly compute $\partial K/ \partial \phi$
  but only $w^T \partial K / \partial \phi$ for the appropriate cotangent vector.
  This type of reasoning plays a key role when
   differentiating functionals of implicit functions -- for example,
   probability densities that depend on solutions to ordinary differential equations --
   and leads to so-called \textit{adjoint methods}
   \citep[e.g. ][]{Errico:1997}.
  
  \subsection{Derivation of the adjoint method}
  
  In this section we provide a proof of theorem~\ref{thm:adjoint}.
  As a starting point, assume algorithm~\ref{algo:diff} is valid.
  The proof can be found in \citet[chapter~5]{Rasmussen:2006}.
  The key observation is that all operations performed on
  $$
    \frac{\partial K}{\partial \phi_j}
  $$
  are linear. Algorithm 1 produces a map
  \begin{eqnarray*}
    \mathcal Z & : \partial K / \partial \phi_j & \to \frac{\partial}{\partial \phi_j} \pi(y \mid \phi) \\
                     & : \mathbb R^{n \times n} & \to \mathbb R,
  \end{eqnarray*}
  and constructs the gradient one element at a time.
  By linearity,
  \begin{eqnarray*} 
      \frac{\partial}{\partial \phi_j} \mathcal Z(K) = 
        \mathcal Z \left ( \frac{\partial K}{\partial \phi_j} \right ).
  \end{eqnarray*}
  Thus an alternative approach to compute the gradient is to calculate 
  the scalar $\mathcal Z(K)$ and then use a single reverse mode sweep
  of automatic differentiation, noting that $\mathcal Z$ is an analytical function.
  This produces Algorithm~\ref{algo:diff3}.
  \begin{algorithm}
    \caption{\textit{Gradient of the approximate marginal log density, 
    $\log \pi_\mathcal{G}(y \mid \phi)$,
    with respect to the hyperparameters, $\phi$}, using reverse mode automatic differentiation}
    \label{algo:diff3}
    \begin{algorithmic}[2]
    \item \textbf{input:} $y$, $\phi$, $\pi(y \mid \theta, \phi)$
    \item Do lines 2 - 6 of Algorithm 2.
    \item Initiate an expression tree for automatic differentiation with $\phi_v = \phi$.
    \item $K_v = \mathcal K(\phi_v)$
    \item $z = \mathcal Z(K_v)$
    \item Do a reverse-sweep over $z$ to obtain $\nabla_\phi \log \pi(y \mid \phi)$.
    \item \textbf{return:} $\nabla_\phi \log \pi(y \mid \phi)$.
    \end{algorithmic}
  \end{algorithm}
  At this point, the most important is done in order to achieve scalability:
  we no longer explicitly compute $\partial K / \partial \phi$ and are using a single reverse mode sweep.
  
  Automatic differentiation, for all its relatively cheap cost,
  still incurs some overhead cost.
  Hence, where possible, we still want to use analytical results to compute derivatives.
  In particular, we can analytically work out the cotangent
  \begin{equation*}
    w^T := \frac{\partial z}{\partial K}.
  \end{equation*}
  For the following calculations, we use a lower case, $k_{ij}$ and $r_{ij}$,
  to denote the $(ij)^\mathrm{th}$ element respectively of the matrices $K$ and $R$.
  
  Consider
  \begin{equation*}
    \mathcal Z(K) = s_1 + s_2^T s_3,
  \end{equation*}
  where, unlike in Algorithm 1, $s_1$ and $s_3$
  are now computed using $K$, not $\partial K / \partial \phi_j$.
 We have
  \begin{eqnarray*}
    s_1 = \frac{1}{2} a^T K a - \frac{1}{2} \mathrm{tr}(RK).
  \end{eqnarray*}
  Then
  \begin{eqnarray*}
  \frac {\partial}{\partial k_{i'j'}} a^T K a =
    \frac {\partial}{\partial k_{i'j'}} \sum_i \sum_j a_i k_{ij} a_j = a_{i'} a_{j'},
  \end{eqnarray*}
  and
  \begin{eqnarray*}
    \frac {\partial}{\partial k_{i'j'}} \mathrm{tr}(RK) = \frac {\partial}{\partial k_{i'j'}} \sum_l r_{il} k_{li} = r_{j' i'}.
  \end{eqnarray*}
  Thus
  \begin{eqnarray*}
    \frac{\partial s_1}{\partial K} = \frac{1}{2} a a^T - \frac{1}{2} R^T.
  \end{eqnarray*}
  For convenience, denote $l = \nabla_\theta \log \pi(y \mid \theta, \phi)$.
  We then have
  \begin{eqnarray*}
  \begin{aligned}
    b & = K l, \\
    s_3 & = b - \tilde KRb = (I - \tilde KR) b,
  \end{aligned}
  \end{eqnarray*}
  where $\tilde K = K$, but is maintained fixed, meaning we do not propagate
  derivatives through it.
  Let $\tilde A = I - \tilde KR$ and let $\tilde a_{ij}$ denote the $(i, j)^\mathrm{th}$
  element of $\tilde A$.
  Then
  \begin{eqnarray*}
    s_2^T s_3 = \sum_i (s_2)_i \left ( \sum_j \tilde a_{ij} \sum_m k_{jm} l_m \right).
  \end{eqnarray*}
  Thus
  \begin{eqnarray*}
    \frac{\partial }{\partial k_{i'j'}} s_2^T s_3 = \sum_i (s_2)_i \tilde a_{i i'} l_{j'} = l_{j'} \sum_i (s_2)_i \tilde a_{i i'},
  \end{eqnarray*}
  where the sum term is the $(i')^\mathrm{th}$ element of $\tilde A s_2$.
  The above expression then becomes
  \begin{eqnarray*}
    \frac{\partial}{\partial K} s^T_2 s_3 = \tilde A s_2 l^T = s_2 l^T - KRs_2 l^T.
  \end{eqnarray*}
  Combining the derivative for $s_1$ and $s^T_2 s_3$ we obtain
  \begin{equation*}
    w^T = \frac{1}{2} aa^T - \frac{1}{2} R 
      + (s_2 + RK s_2) [\nabla_\theta \log \pi(y \mid \theta, \phi)]^T,
  \end{equation*}
  as prescribed by Theorem~\ref{thm:adjoint}.
  This result is general, in the sense that it applies
  to any covariance matrix, $K$, and likelihood, $\pi(y \mid \theta, \phi)$.
  Our preliminary experiments, on the SKIM, found that incorporating
  the analytical cotangent, $w^T\!$, approximately doubles the differentiation speed.

  \section{Computer code}
  
  The code used in this work is open source and detailed in this section.
  
  \subsection{Prototype Stan code}
  
  The Stan language allows users to specify the joint log density of their model.
  This is done by incrementing the variable \textcolor{magenta}{\texttt{target}}.
  We add a suite of functions, which return the approximate log marginal density, 
  $\log \pi_\mathcal{G}(y \mid \phi)$.
  Hence, the user can specify the log joint distribution by incrementing
  \textcolor{magenta}{\texttt{target}} with $\log \pi_\mathcal{G}(y \mid \phi)$
  and the prior $\log \pi(\phi)$.
  A call to the approximate marginal density looks as follows:
  \begin{lstlisting}[style=stan, numbers=none] 
target +=
  <@\textcolor{codepurple}{laplace\_marginal\_*\_lpmf}@> (y | n, K, phi, x, delta, 
                           delta_int, theta0);
  \end{lstlisting}
  The \textcolor{codepurple}{\texttt{*}} specifies the obervation model, typically a distribution and a link function, 
  for example \textcolor{codepurple}{\texttt{bernoulli\_logit}} or 
  \textcolor{codepurple}{\texttt{poisson\_log}}.
  The suffix \texttt{lpmf} is used in Stan to denote a log posterior mass function.
  \texttt{y} and \texttt{n} are sufficient statistics for the latent Gaussian variable, $\theta$;
  \texttt{K} is a function that takes in arguments \texttt{phi}, \texttt{x}, \texttt{delta}, and \texttt{delta\_int}
  and returns the covariance matrix;
  and \texttt{theta0} is the initial guess for the Newton solver, which seeks the mode of $\pi(\theta \mid \phi, y)$.
  Moreover, we have
  \begin{itemize}
    \item \texttt{y}: a vector containing the sum of counts/successes for each element of $\theta$,
    \item \texttt{n}: a vector with the number of observation for each element of $\theta$,
    \item \texttt{K}: a function defined in the functions block, with the signature
    \texttt{(vector, data matrix, data real[], data int[]) ==> matrix}. 
    Note that only the first argument may be used to pass variables which depend on model parameters,
    and through which we propagate derivatives.
    The term \texttt{data} means an argument may not depend on model parameters.
    \item \texttt{phi}: the vector of hyperparameters,
    \item \texttt{x}: a matrix of data. For Gaussian processes, this is the coordinates,
    and for the general linear regression, the design matrix,
    \item \texttt{delta}: additional real data,
    \item \texttt{delta\_int}: additional integer data,
    \item \texttt{theta0}: a vector of initial guess for the Newton solver.
  \end{itemize}
  It is also possible to specify the tolerance of the Newton solver.
  This structure is consistent with other higher-order functions in Stan,
  such as the algebraic solver and the ordinary differential equation solvers.
  It gives users  flexibility when specifying $K$, but we recognize it is cumbersome.
  One item on our to-do list is to use variadic arguments,
  which remove the constraints on the signature of \texttt{K},
  and allows users to pass any combination of arguments to \texttt{K}
  through \texttt{laplace\_marignal\_*\_lpmf}.

  For each observation model, we implement 
  a corresponding random number generating function,
  with a call
    \begin{lstlisting}[style=stan, numbers=none] 
theta =  <@\textcolor{codepurple}{laplace\_marginal\_*\_rng}@> (y, n, K, phi, x, delta, 
                             delta_int, theta0);
  \end{lstlisting}

  This generates a random sample from $\pi_\mathcal{G}(\theta \mid y, \phi)$.
  This function can be used in the generated quantities blocks
  and is called only once per iteration --
  in contrast with the target function which is called and differentiated
  once per integration step of HMC.
  Moreover the cost of generating $\theta$ is negligible next to the
  cost evaluating and differentiating $\log \pi(y \mid \phi)$
  multiple times per iteration.
  
  The interested reader may find a notebook with demo code, including R scripts and Stan files,
  at \url{https://github.com/charlesm93/StanCon2020},
  as part of the 2020 Stan Conference \cite{Margossian:2020}.

  \subsection{C++ code}
  
  We incorporate the Laplace suite of functions inside the Stan-math
  library, a C++ library for automatic differentiation \citep{Carpenter:2015}.
  The library is open source and available on GitHub, \url{https://github.com/stan-dev/math}.
  Our most recent prototype exists on the branch \texttt{try-laplace\_approximation2}\footnote{
  Our first prototype is was on the branch \texttt{try-laplace\_approximation}, and was used to
  conduct the here presented computer experiment. 
  The new branch modifies the functions' signatures to be more consistent with the Stan language.
  In this Supplement, we present the new signatures.}.
  The code is structured around a main function
\begin{lstlisting}[style=stan, numbers=none]
  <@\textcolor{codepurple}{laplace\_approximation}@> (likelihood, K_functor, phi, x, delta, 
                      delta_int, theta0);
\end{lstlisting}
  with
  \begin{itemize}
    \item \texttt{likelihood:} a class constructed using \texttt{y} and \texttt{n},
      which returns the log density, as well as its first, second, and third order derivatives.
    \item \texttt{K\_functor:} a functor that computes the covariance matrix, $K$
    \item ...: the remaining arguments are as previously described. 
  \end{itemize}
  A user can specify a new likelihood by creating the corresponding class,
  meaning the C++ code is expandable.
  
  To expose the code to the Stan language,
  we use Stan's new OCaml transpiler, stanc3, \url{https://github.com/stan-dev/stanc3}
  and again the branch \texttt{try-laplace\_approximation2}.
  
  Important note: the code is prototypical and currently not merged into Stan's release
  or development branch.
  
  \subsection{Code for the computer experiment}
  
  The code is available on the GitHub public repository,
  \url{https://github.com/charlems93/laplace\_manuscript}.
  
  We make use of two new prototype packages:
  \mbox{CmdStanR} (\url{https://mc-stan.org/cmdstanr/})
  and posterior (\url{https://github.com/jgabry/posterior}).

  \section{Tuning dynamic Hamiltonian Monte Carlo}

  In this article, we use the dynamic Hamiltonian Monte Carlo sampler described 
  by \citet{Betancourt:2018} and implemented in Stan.
  This algorithm builds on the No-U Turn Sampler by \citet{Hoffman:2014},
  which adaptively tunes the sampler during a warmup phase.
  Hence for most problems, the user does not need to worry about tuning 
  parameters.
  However, the models presented in this article are challenging
  and the sampler requires careful tuning,
  if we do not use the embedded Laplace approximation.
  
  The main parameter we tweak is the \textit{target acceptance rate}, $\delta_a$.
  To run HMC, we need to numerically compute physical trajectories
  across the parameter space
  by solving the system of differential equations prescribed by Hamilton's equations
  of motion.
  We do this using a numerical integrator.
  A small step size, $\delta$, makes the integrator more precise
  but generates smaller trajectories, which leads to a less efficient exploration
  of the parameter space.
  When we introduce too much numerical error, the proposed trajectory is rejected.
  Adapt delta, $\delta_a \in (0, 1)$, sets the target acceptance rate of proposed trajectories.
  During the warmup, the sampler adjusts $\delta$ to meet this target.
  For well-behaved problems, the optimal value of $\delta_a$ is 0.8 \cite{Betancourt:2015}.
  
  It should be noted that the algorithm does not necessarily achieve the target set by $\delta_a$
  during the warmup.
  One approach to remedy this issue is to extend the warmup phase;
  specifically the final fast adaptation interval or \textit{term buffer}
  \citep[see ][]{Hoffman:2014, Stan:2020}.
  By default, the term buffer runs for 50 iterations (when running a warmup for 1,000 iterations).
  Still, making the term buffer longer does not guarantee the sampler attains the target $\delta_a$.
  There exist other ways of tuning the algorithm,
  but at this points, the technical burden on the user is already significant.
  What is more, probing how well the tuning parameters work usually requires running the model
  for many iterations.
  
  \section{Automatic differentiation variational inference}
  
  ADVI automatically derives a variational inference algorithm, based on a user specified log joint density.
  Hence we can use the same Stan file we used for full HMC and, with the appropriate call,
  run ADVI instead of MCMC.
  The idea behind ADVI is to approximate the posterior over the unconstrained space using a Gaussian distribution,
  either with a diagonal covariance matrix -- leading to a mean-field approximation -- 
  or with a full rank covariance matrix.
  The details of this procedure are described in \cite{Kucukelbir:2017}.
  Compared to full HMC, ADVI can be much faster, 
  but in general it is difficult to assess how well the variational approximation describes 
  the target posterior distribution without using an expensive benchmark \cite{Yao:2018, Huggins:2020}.
  Furthermore, it can be challenging to assess the convergence of ADVI \cite{Dhaka:2020}.

  To run ADVI, we use the Stan file with which we ran full HMC.
  We depart from the default tuning parameters by decreasing the learning rate $\eta$ to 0.1,
  adjusting the tolerance, \texttt{rel\_tol\_obj},
  and increasing the maximum number of iterations to 100,000.
  Our goal is to improve the accuracy of the optimizer as much as possible, while insuring that convergence is reached.
  
\begin{figure}
    \includegraphics[width=5.5in]{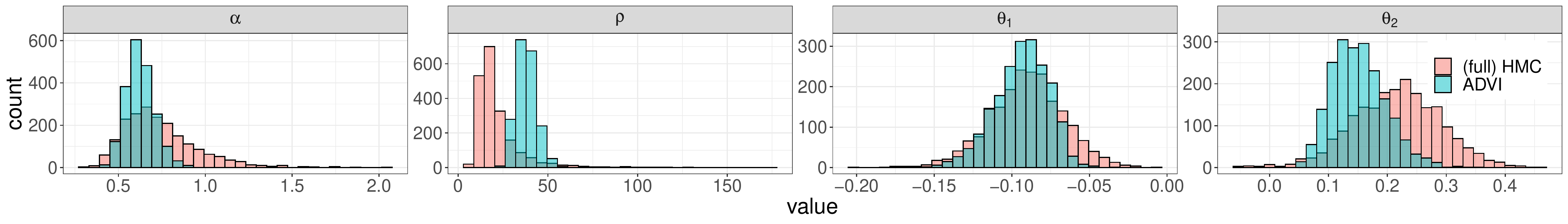}
    \caption{Samples obtained with full HMC and sampling from the variational approximation produced by ADVI
    when fitting the disease map. 
    Unlike the embedded Laplace approximation, ADVI strongly disagrees with full HMC.}
    \label{fig:gp_advi_sample}
  \end{figure}

\begin{figure}
    \includegraphics[width=5.5in]{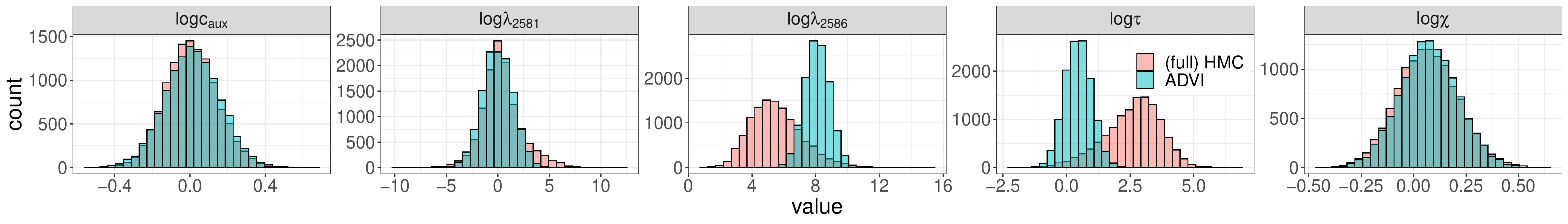}
    \caption{Samples obtained with full HMC and sampling from the variational approximation produced by ADVI
    when fitting the SKIM.}
    \label{fig:skim_advi_sample}
  \end{figure}

  We compare the samples drawn from the variational approximation
  to samples drawn from full HMC in Figures~\ref{fig:gp_advi_sample}, \ref{fig:horseshoe_advi_sample}
  and \ref{fig:skim_advi_sample}. 
  For the studied examples, we find the approximation to be not very satisfactory,
  either because it underestimates the posterior variance,
  does not capture the skewness of the posterior distribution,
  or returns a unimodal approximation when in fact the posterior density is multimodal.
  These are all features which cannot be captured by a Gaussian over the unconstrained scale.
  Naturally, a different choice for $\mathcal Q$ could lead to better inference.
  Using a custom VI algorithm is however challenging, as we need to derive a useful variational family
  and hand-code the inference algorithm, rather than rely on the implementation in
  a probabilistic programming language.

  
  \section{Model details}
  
  We review the models used in our computer experiments
  and point the readers to the relevant references.
  
  \subsection{Disease map}
  
  The disease map uses a Gaussian process with an exponentiated squared kernel,
  \begin{eqnarray*}
     k(x_i, x_j) = \alpha^2 \exp \left (- \frac{(x_i - x_j)^T(x_i - x_j)}{\rho^2} \right).
  \end{eqnarray*}
  The full latent Gaussian model is
  \begin{eqnarray*}
      \rho & \sim & \mathrm{invGamma}(a_\rho, b_\rho), \\
      \alpha & \sim & \mathrm{invGamma}(a_\alpha, b_\alpha), \\
      \theta & \sim & \Normal(0, K(\alpha, \rho, x)), \\
      y_i & \sim & \mathrm{Poisson}(y_e^i e^{\theta_i}),
  \end{eqnarray*}
  where we put an inverse-Gamma prior on $\rho$ and $\alpha$.

  When using full HMC, we construct a Markov chain over the joint parameter space $(\alpha, \rho, \theta)$.
  To avoid Neal's infamous funnel \cite{Neal:2003}
  and improve the geometry of the posterior distribution,
  it is possible to use a \textit{non-centered parameterization}:
  \begin{eqnarray*}
    (\rho, \alpha) & \sim & \pi(\rho, \alpha), \\
    z & \sim & \Normal(0, I_{n \times n}), \\
    L & = & \mathrm{Cholesky \ decompose}(K), \\
    \theta & = & L z, \\
    y_i & \sim & \mathrm{Poisson}(y^i_e e^{\theta_i}).
  \end{eqnarray*}
  The Markov chain now explores the joint space of $(\alpha, \rho, z)$
  and the $\theta$'s are generated by transforming the $z$'s.  
  With the embedded Laplace approximation, the Markov chain
  only explores the joint space $(\alpha, \rho)$.
  
  To run ADVI, we use the same Stan file as for full HMC and set \texttt{tol\_rel\_obj} to 0.005.
  
  \subsection{Regularized horseshoe prior} \label{app:reg_horse_prior}
  
  The horseshoe prior \citep{Carvalho+Polson+Scott:2010:HS} 
  is a sparsity inducing prior that introduces a global shrinkage parameter, $\tau$,
  and a local shrinkage parameter, $\lambda_i$ for each covariate slope, $\beta_i$.
  This prior operates a soft variable selection, 
  effectively favoring $\beta_i \approx 0$ or $\beta_i \approx \hat \beta_i$,
  where $\hat \beta_i$ is the maximum likelihood estimator.
  \citet{Piironen:2017} add another prior to regularize unshrunk $\beta$s, 
  $\Normal(0, c^2)$, effectively operating a ``soft-truncation''
  of the extreme tails.

  \subsubsection{Details on the prior}
  
  For computational stability, the model is parameterized using $c_\mathrm{aux}$,
  rather than $c$, where
  \begin{equation*}
      c = s_\mathrm{slab} \sqrt{c_\mathrm{aux}}
  \end{equation*}
  with $s_\mathrm{slab}$ the slab scale.
  The hyperparameter is $\phi = (\tau, c_\mathrm{aux}, \lambda)$
  and the prior
  \begin{eqnarray*}
     \lambda_i & \sim & \mathrm{Student}_t(\nu_\mathrm{local}, 0, 1), \\
     \tau & \sim &  \mathrm{Student}_t(\nu_\mathrm{global}, 0, s_\mathrm{global}),  \\
     c_\mathrm{aux} & \sim & \mathrm{inv}\Gamma(s_\mathrm{df} / 2, s_\mathrm{df} / 2), \\
     \beta_0 & \sim & \Normal(0, c_0^2).
  \end{eqnarray*}
  The prior on $\lambda$ independently applies to each element, $\lambda_i$.
  
  Following the recommendation by \citet{Piironen:2017},
  we set the variables of the priors as follows.
  Let $p$ be the number of covariates and $n$ the number of observations.
  Additionally, let $p_0$ be the expected number of relevant covariates --
  note this number does not strictly enforce the number of unregularized $\beta$s,
  because the priors have heavy enough tails that we can depart from $p_0$.
  For the prostate data, we set $p_0 = 5$.
  Then
  \begin{eqnarray*}
    s_\mathrm{global} & = & \frac{p_0}{\sqrt{n}(p - p_0)}, \\
    \nu_\mathrm{local} & = & 1, \\
    \nu_\mathrm{global} & = & 1, \\
    s_\mathrm{slab} & = & 2, \\
    s_\mathrm{df} & = & 100, \\
    c_0 & = & 5.
  \end{eqnarray*}
  Next we construct the prior on $\beta$,
  \begin{equation*}
    \beta_i \sim \Normal(0, \tau^2 \tilde \lambda_i^2),
  \end{equation*}
  where
  \begin{equation*}
    \tilde \lambda_i^2 =  \frac{c^2 \lambda_i^2}{c^2 + \tau^2 \lambda_i^2}.
  \end{equation*}

  \subsubsection{Formulations of the data generating process}
  
  The data generating process is
    \begin{eqnarray*}
    \phi & \sim & \pi(\phi),  \\
    \beta_0 & \sim & \Normal(0, c_0^2),  \\
    \beta & \sim & \Normal(0, \Sigma(\phi)), \\
    y & \sim & \mathrm{Bernoulli\_logit}(\beta_0 + X \beta),
  \end{eqnarray*}
  or, equivalently,
  \begin{eqnarray*}
    \phi & \sim & \pi(\phi), \\
    \theta & \sim & \Normal(0,\, c_0^2 I_{n \times n} + X \Sigma(\phi) X^T),  \\
    y & \sim & \mathrm{Bernoulli\_logit}(\theta).
  \end{eqnarray*}
  
  For full HMC, we use a non-centered parameterization of the first formulation,
  much like we did for the disease map.
  The embedded Laplace approximation, as currently implemented, requires the second formulation,
  which is mathematically more convenient but 
  comes at the cost of evaluating and differentiating
  \mbox{$K = c^2 I_{n \times n} + X \Sigma(\phi) X^T$}.
  In this scenario, the main benefit of the Laplace approximation 
  is not an immediate speed-up
  but an improved posterior geometry, due to marginalizing $\theta$
  (and thus implicitly $\beta$ and $\beta_0$) out.
  This means we do not need to fine tune the sampler.
  
  \subsubsection{Fitting the model with full HMC}
  
   This section describes how to tune full HMC to fit the model at hand.
   Some of the details may be cumbersome to the reader.
   But the takeaway is simple: tuning the algorithm is hard
   and can be a real burden for the modeler.
   
   Using a non-centered parameterization and
   with Stan's default parameters, we obtain $\sim$150 divergent transitions\footnote{
   To be precise, we here did a preliminary run using 4000 sampling iterations 
   and obtained 50 divergent transitions
   (so an expected 150 over 12000 sampling iterations).
   }.
   We increase the target acceptance rate to $\delta_a = 0.99$
   but find the sampler now produces 186 divergent transitions.
   A closer inspection reveals the divergences all come from a single chain,
   which also has a larger adapted step size, $\delta$.
   The problematic chain also fails to achieve the target acceptance rate.
   These results are shown in Table~\ref{tab:adaptive}.
   From this, it seems increasing $\delta_a$ yet again may not provide any benefits.
   Instead we increase the term buffer from 50 iterations to 350 iterations.
   With this setup, we however obtain divergent transitions across all chains.
  
 \begin{table}
    \caption{Adapted tuning parameters across 4 Markov chains with $\delta_a = 0.99$.}
   \label{tab:adaptive}
   \renewcommand{\arraystretch}{1.5}
   \begin{center}
   \begin{tabular}{c r r r}
     \rowcolor[gray]{0.95} \textbf{Chain} & \textbf{Step size} & \textbf{Acceptance rate} & 
     \textbf{Divergences} \\
     1 & 0.0065  & 0.99 & 0 \\
     \rowcolor[gray]{0.95}  \textcolor{red}{2} & \textcolor{red}{0.0084} &
      \textcolor{red}{0.90} & \textcolor{red}{186} \\
     3 & 0.0052 & 0.99 & 0 \\
     \rowcolor[gray]{0.95} 4 & 0.0061 & 0.99 & 0
   \end{tabular}
   \end{center}
 \end{table}
 
  This outcome indicates the chains are relatively unstable and emphasizes how difficult it is,
  for this type of model and data, to come up with the right tuning parameters.
  With $\delta_a = 0.999$ and the extended term buffer we observe
  13 divergent transitions.
  It is possible this result is the product of luck, rather than better tuning parameters.
  To be clear, we do not claim we found the optimal model parameterization
  and tuning parameters.
  There is however, to our knowledge, no straightforward way to do so.

  \subsubsection{Fitting the model with the embedded Laplace approximation}
  
  Running the algorithm with Stan's default tuning parameters produces 0 divergent transitions
  over 12,000 sampling iterations.
  
  \subsubsection{Fitting the model with ADVI}
  
  To run ADVI, we use the same Stan file as for full HMC and set \texttt{tol\_rel\_obj} to 0.005.
  
\begin{figure}
    \includegraphics[width=5.5in]{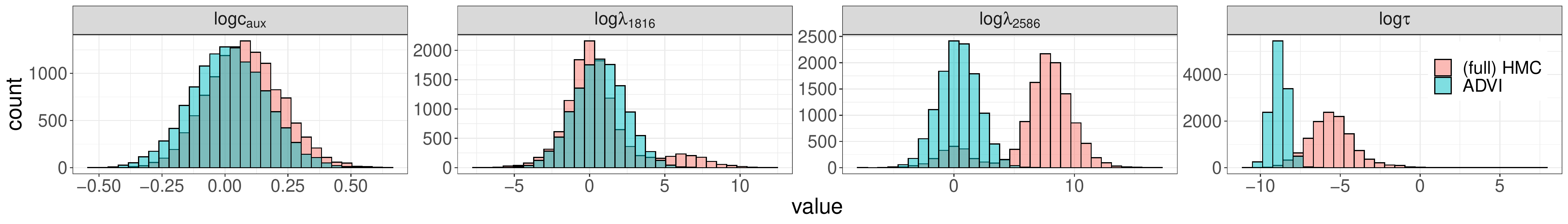}
    \caption{Samples obtained with full HMC and sampling from the variational approximation produced by ADVI
    when fitting a general linear model with a regularized horseshoe prior.}
    \label{fig:horseshoe_advi_sample}
  \end{figure}

  The family of distribution, $\mathcal Q$, over which ADVI optimizes requires the exact posterior distribution
  to be unimodal over the unconstrained scale.
  This is a crucial limitation in the studied example, as shown in Figure~\ref{fig:horseshoe_advi_sample}.
  This notably affects our ability to select relevant covariates using the $90^\mathrm{th}$ posterior quantile.
  When examining the top six selected covariates (Table 1 in the main text), we find the result from ADVI
  to be in disagreement with full HMC and the embedded Laplace approximation.
  In particular, $\lambda_{2586}$ which corresponds, according to our other inference methods,
  to the most relevant covariate, has a relatively low $90^\mathrm{th}$ quantile.
  This is because ADVI only approximates the smaller mode of $\pi(\lambda_{2586} \mid y)$.
  Our results are consistent with the work by \citet{Yao:2018}, who examine ADVI on a similar problem.
  
  \subsection{Sparse kernel interaction model}
  \label{app:skim}
      
  SKIM, developed by \citet{Agrawal:2019}, 
  extends the model of \citet{Piironen:2017} by accounting
  for pairwise interaction effects between covariates. The generative model shown below uses the notation in \ref{app:reg_horse_prior} instead of that in Appendix D of \citet{Agrawal:2019}:
\begin{align*}
\begin{split}
  \chi &\sim \mathrm{inv}\Gamma(s_\mathrm{df} / 2, s_\mathrm{df} / 2), \\
 \eta_2 &= \frac{\tau^2}{c^2} \chi, \\
\beta_{i} \mid \tau, \tilde{\lambda} &\sim \Normal(0, \tau^2 \tilde{\lambda}_i^2),  \\
\beta_{j} \mid \tau, \tilde{\lambda} &\sim \Normal(0, \tau^2 \tilde{\lambda}_i^2),  \\
\beta_{ij} \mid \eta_2, \tilde{\lambda} &\sim \Normal(0, \eta_2^2 \tilde{\lambda}_i^2 \tilde{\lambda}_j^2), \\
\beta_0 \mid c_0^2 &\sim \Normal(0, c_0^2),
\end{split}
\end{align*}
  where $\beta_i$ and $\beta_{ij}$ are the main and pairwise effects for covariates $x_i$ and $x_ix_j$, respectively, and $\tau$, $\tilde{\lambda}$, $c_0$ are defined in \ref{app:reg_horse_prior}.  
  
Instead of sampling $\{\beta_i \}_{i=1}^p$ and  $\{\beta_{ij} \}_{i,j=1}^p$, which takes at least $O(p^2)$ time per iteration to store and compute, \citet{Agrawal:2019} marginalize out all the regression coefficients, only sampling $(\tau, \xi, \tilde{\lambda})$ via MCMC. Through a kernel trick and a Gaussian process re-parameterization of the model, this marginalization takes $O(p)$ time instead of $O(p^2)$.  The Gaussian process covariance matrix $K$ induced by SKIM is provided below:

  \begin{eqnarray*}
    K_1 & = & x \ \mathrm{diag}(\tilde{\lambda}^2) \ x^T, \\
    K_2 & = & [x \circ x] \ \mathrm{diag}(\tilde{\lambda}^2) \ [x \circ x]^T,
  \end{eqnarray*} 
   where ``$\circ$'' denotes the element-wise Hadamard product.
   Finally,
   \begin{eqnarray*}
    K & = & \frac{1}{2} \eta_2^2 (K_1 + 1) \circ (K_1 + 1) - \frac{1}{2} \eta_2^2 K_2
    - (\tau^2 - \eta_2^2) K_1 \\ 
    & & + c_0^2  - \frac{1}{2} \eta_2^2.
  \end{eqnarray*}
  
  \bibliography{ref_laplace.bib}
  \bibliographystyle{abbrvnat}

\end{document}